\documentclass[conference,a4paper]{IEEEtran}
\usepackage{amsmath,amsfonts}
\usepackage{array}
\usepackage[caption=false,font=normalsize,labelfont=sf,textfont=sf]{subfig}
\usepackage{textcomp}
\usepackage{stfloats}
\usepackage{url}
\usepackage{verbatim}
\usepackage{graphicx}
\usepackage{cite}
\usepackage[linesnumbered,ruled,vlined]{algorithm2e}

\begin{document}

\title{BeerReview: A Blockchain-enabled Peer Review Platform}

\author{\IEEEauthorblockN{Guodong Jin}
\IEEEauthorblockA{\textit{College of Electronic and Information }\\ \textit{  Engineering} \\
\textit{Tongji University}\\
2252540@tongji.edu.cn}
\and

\IEEEauthorblockN{Zihan Zhou}
\IEEEauthorblockA{\textit{Information Hub of HKUST} \\
\textit{Hong Kong University of Science }\\
\textit{and Technology (Guangzhou)} \\
zzh45472@gmail.com}
\and

\IEEEauthorblockN{Wenzheng Tang}
\IEEEauthorblockA{\textit{College of Electronic and Information }\\ \textit{  Engineering} \\
\textit{Tongji University}\\
2232962@tongji.edu.cn}
\and

\IEEEauthorblockN{Kanglei Yu}
\IEEEauthorblockA{\textit{College of Electronic and Information }\\ \textit{  Engineering} \\
\textit{Tongji University}\\
2152206@tongji.edu.cn}
\and

\IEEEauthorblockN{Hao Xu}
\IEEEauthorblockA{\textit{College of Electronic and Information }\\ \textit{  Engineering} \\
\textit{and SERCBAAS}\\
\textit{Tongji University}\\
Hao.Xu@ieee.org}

\and
\IEEEauthorblockN{Erwu Liu}
\IEEEauthorblockA{\textit{College of Electronic and Information }\\ \textit{  Engineering} \\
\textit{and SERCBAAS}\\
\textit{Tongji University}\\
erwuliu@tongji.edu.cn}
}



\maketitle

\begin{abstract}
In an era of increasing concerns over intellectual property rights, traditional peer review systems face challenges including plagiarism, malicious attacks, and unauthorized data access. BeerReview, a blockchain-enabled peer review platform, offers a robust solution, enabling experts and scholars to participate actively in the review process without concerns about plagiarism or security threats. Following the completion of its alpha testing, BeerReview demonstrates the potential for expanded deployment. This platform offers improved convenience and more robust intellectual property protection within the peer review process with open source initiative.
\footnote{Project repository: \url{https://github.com/Tongji-Blockchain}}
\end{abstract}


\section{Introduction}

\IEEEPARstart{P}{eer} review, defined as “a process of subjecting an author’s scholarly work, research or ideas to the scrutiny of others who are experts in the same field” \cite{sowards2015peer}, stands as an indispensable process along the scientific journey of every researcher. The peer review process serves as a critical filter, verifying the validity, significance, and originality of research before publication. Additionally, it provides constructive feedback that prompts authors to refine their manuscripts to align with the rigorous standards requisite for acceptance in their respective disciplines \cite{kelly2014}.

However, in an era marked by increasing concerns over intellectual property rights (IPRs) and the fluid nature of collaborative research, traditional peer review systems confront a myriad of challenges, which extend beyond mere evaluation to encompass the protection of authors' intellectual contributions, raising questions about integrity, confidentiality, attribution, and ownership \cite{kelly2014}. There are mainly three ways to conduct the peer review process, all of which have disadvantages: a)open review, b)single-blind review, and c)double-blind review \cite{kelly2014}.
\begin{itemize}
    \item \textbf{Open peer review} can prevent reviewers from being honest, since they may worry about the breakdown of their relationship with the author. Out of politeness, they would express their criticisms in a more discreet manner \cite{elsevier2014}.
    \item \textbf{Single-blind peer review} conceals the reviewers' identity to promote honesty, but the reviewers who receive manuscripts on subjects similar to their own research may delay completing the review process deliberately to publish their own papers first \cite{elsevier2014}. 
    \item \textbf{Double-blind peer review} seems to be the most popular way of peer review, as The Sense About Science survey indicates that 76\% of researchers praise it \cite{schley2009}. However, a shortcoming is that reviewers can easily determine the identity of the author by the writing style, subject matter, or self-citation, potentially introducing bias \cite{elsevier2014}.
\end{itemize}

Moreover, the platforms used for peer review also have potential threats. Traditional peer review systems are based on a centralized platform, which is vulnerable to malicious attacks that endanger users’ IPRs and is unable to guarantee the authenticity of information loggings \cite{syed2020, kulkarni2012}. Even well-known system platforms, such as Magtech Editorial System \cite{magtech} and Samson Editorial System \cite{samsoncn}, carry underlying risks of malicious attacks and data tampering. 

A decentralized platform can effectively address this issue. Here are the reasons.
\begin{itemize}
    \item \textbf{Security of IPRs} Although in a centralized platform, controlling who can access the information is feasible, malicious attacks against the central administrator, who can get access to all the information, still pose a threat to users' IPRs. However, in a decentralized platform, without the central administrator, it becomes impractical to compromise the entire system by attacking only a few nodes, in other words, the cost of attacking the system is extremely high.
    \item \textbf{Authenticity of Information Loggings} The distributed consensus protocol is the key technology that enables blockchain’s decentralization, or more specifically, that ensures all participants agree on a unified transaction ledger without the management of a central authority \cite{xiao2020}. Therefore, attackers can't intrude into the system without violating the consensus protocol, which guarantees the authenticity of information loggings stored in the decentralized system.
\end{itemize}

Blockchain technology \cite{buterin2014}, which drives an information revolution, further impacting society and social sciences \cite{li2022recordism}, is an excellent example of decentralized methods.
Following a review of existing data sharing systems and frameworks based on blockchain technology \cite{guo2022, klaine2023}, this paper introduces BeerReview, a blockchain-enabled peer review platform. A detailed introduction to this platform will be provided, highlighting its capabilities to address the identified challenges in traditional peer review systems. The solutions to these challenges are outlined in the subsequent sections.

The study presents specific solutions to address the challenges within the three traditional methods of conducting the peer review process:
\begin{itemize}
    \item For the problem of open review, in BeerReview, users can utilize aliases, initials, or even names unrelated to their identities as their usernames so that they can follow their heart to review the manuscripts.
    \item A manuscript in the peer review process conducted on BeerReview will pass the peer review process once it receives a sufficient number of favorable opinions while reaching the preset percentage of total reviewers, which can handle deliberate delay effectively.
    \item A large language model (LLM) instance is applied on BeerReview to generate abstracts for the Articles, which conceals details that may indicate the author's identity.
\end{itemize}

\begin{figure}
    \centering
    \includegraphics[width=0.48\textwidth]{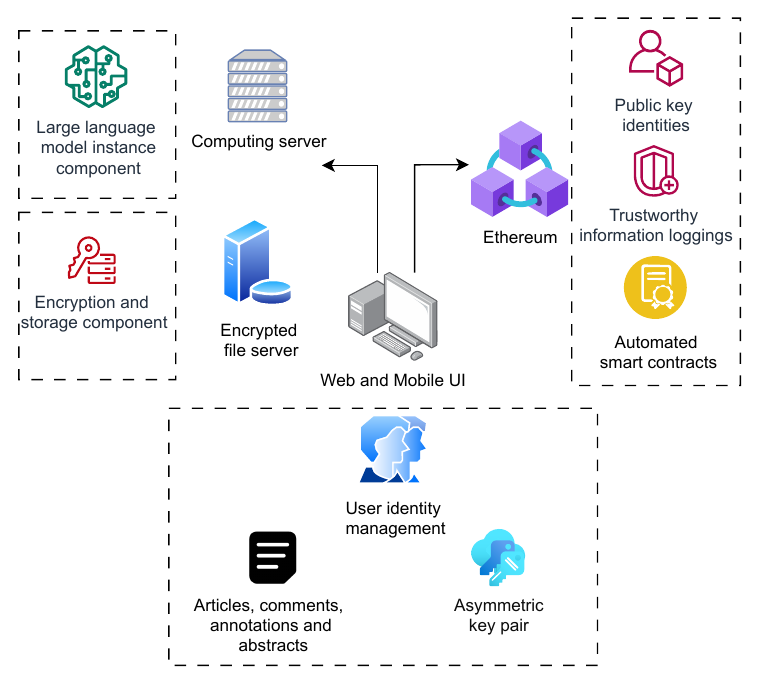}
    \caption{The overall framework of BeerReview.}
    \label{fig: Framework}
\end{figure}

To mitigate the potential risks associated with centralized peer review platforms, this study advocates the adoption of Ethereum, a decentralized platform that enables secure and transparent transactions.

\begin{itemize}
    \item \textbf{Structure} Ethereum, considered as a whole, functions as a transaction-based state machine starting from a genesis state and incrementally executing transactions to evolve into its current state \cite{wood2022}. Transactions are collated into blocks, which are chained together using a cryptographic hash as a means of reference \cite{wood2022}.
    \item \textbf{Features} On Ethereum, all parties are only allowed to create a new block on older pre-existing blocks, after which they can't make any modifications to the block. The two operations parties can do are creating new blocks and reading the information stored on blocks \cite{wood2022, huang2014}.
\end{itemize}

The structure and features of Ethereum make itself a remarkable decentralized tool, which not only guarantees the security of the data stored on Ethereum, preventing them from malicious attacks but also ensures data integrity, making the records of transactions reliable. 

Next, the design and implementation details of BeerReview will be described.

\section{System Design}

\subsection{Framework}

\begin{figure*}
    \centering
    \includegraphics[width=0.6\textwidth,height=9cm]{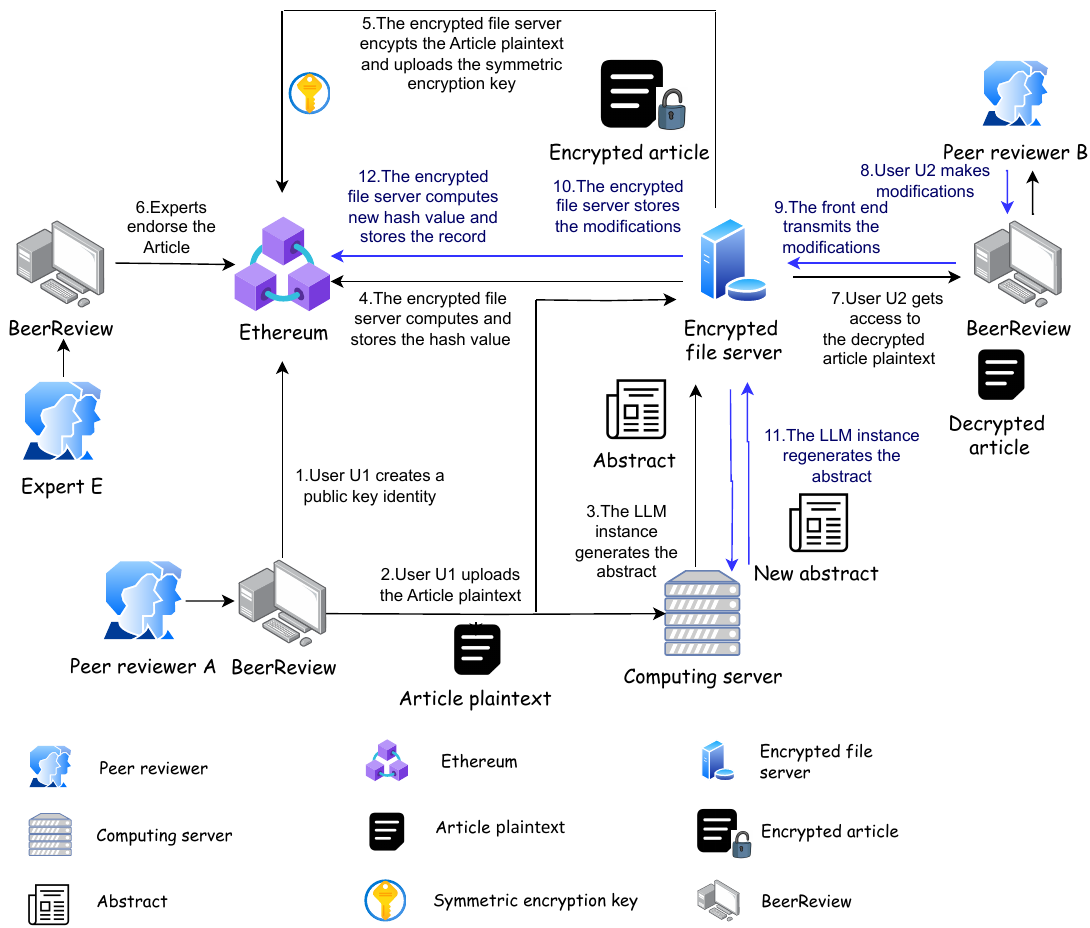}
    \caption{The usage process of BeerReview.}
    \label{fig: Usage Process}
\end{figure*}

The overall framework for BeerReview is shown in Fig. 1. To provide scholars and experts access to the blockchain network, a server works as the encrypted file server is built, through which scholars and experts can log in, access Articles, and make modifications to Articles with their personal computers. The platform consists of three main parts: 1) the Web and Mobile User Interface (UI), 2)the encrypted file server and computing server, and 3)Ethereum. 
\subsubsection{The Web and Mobile UI}
the Web and Mobile UI works as the communicator between the network system and the user. Users can register, read, and comment on the Articles through the UI, and the data from the back end will also be shown to the users on it. The web UI of BeerReview is shown in Fig. 2.
\subsubsection{The Encrypted File Server and Computing Server}
the encrypted file server and computing server have two vital components: 1) the encryption and storage component and 2) the LLM instance component. The encryption storage component realizes file encryption using the asymmetric key pair for the corresponding authority group, while the LLM instance component provides abstracts for given Articles. 
\subsubsection{Ethereum}
Ethereum has multiple capabilities, including creating and managing public key identities, storing trustworthy information loggings, and interacting with the automated smart contract.

\subsection{Usage Process}
For uploaders, as indicated by the black words in Fig. 2, seven steps are required to work with BeerReview. First of all, in step 1, when scholars or experts register for their accounts, their public key identities and blockchain wallets are created. At the same time, the user’s asymmetric key pair is generated using the SM2 algorithm. Then, in step 2, scholars or experts can upload their Articles, which will be transmitted to both the computing server and the encrypted file server in plaintext. After that, in step 3, the LLM instance will generate an abstract for the plaintext of the Article under the drive of the computing server. Then, in step 4, the encrypted file server will compute the hash value of the original plaintext of the Article and store it on Ethereum to prevent tampering. Next, in step 5, the encrypted file server will encrypt the Article using the corresponding asymmetric key pair, and store the symmetric key file encrypted with the asymmetric key and the asymmetric key on Ethereum to enhance the security of the Article. Furthermore, in step 6, authorized peers from the same industry will endorse the Article based on the corresponding article digest generated in step 4. Once the Article is endorsed by experts, in step 7, other scholars and experts within the corresponding authority group can access the article, including reading the full text, posting comments, and making annotations.

For readers, as indicated by the blue words in Fig. 2, the process will be more streamlined. Besides the step 1, five steps should be taken to interact with BeerReview. In steps 8 and 9, a user makes modifications to the Article, which will be transmitted to the encrypted file server by the front end. Next, in step 10, the encrypted file server will store the modifications. After that, in step 11, the new Article will be transmitted to the computing server by the encrypted file server so that the computing server can drive the LLM instance to regenerate a new abstract. Finally, in step 12, the encrypted file server will store the new Article abstract, compute the new hash value of the new Article plaintext, and store the record including the modifier, the modification time, the modified Article ID, and the new hash value.

\subsection{Trustworthy Server}
A trustworthy server is vital for safeguarding user identities and IPRs, incorporating two main elements: 1) P2P encrypted communication and 2) trustworthy services, enhanced by Trusted Execution Environment (TEE) \cite{TEE} and Trusted Platform Module (TPM) \cite{module2018trusted}.

\subsubsection{P2P Encrypted Communication} 
considering the interactions between users and the two servers on BeerReview are similar, the realizations of the identity-encrypted communication for the two servers can be identical.

BeMutual, which is a blockchain-enabled system of
identity management and mutual authentication protocol \cite{zhou2023be}, should be applied to BeerReview as the core technology for P2P encrypted communication. Based on the decentralized public key identity generated for each user, BeMutual can make P2P communication reliable \cite{zhou2023be}. 

\subsubsection{Trustworthy Services} the trustworthy services offered by BeerReview's servers vary based on their specific roles:

\begin{itemize}
    \item \textbf{The Encrypted File Server} The hash value of the Article uploaded by users and the loggings regarding any modifications from other reviewers should be transmitted and stored on Ethereum, which ensures the data integrity. Utilizing TPM ensures the integrity and security of data storage and access, making unauthorized data modification detectable and preventable. Additionally, cryptographic operations, including the encryption and decryption of data, are managed by user-specific keys held within TPM, enhancing security and ensuring that only authorized users can access or modify their data.
    \item \textbf{The Computing Server} Smart Contract-based Computing Server Cluster Consensus Protocol can be adopted by BeerReview to realize trustworthy abstracts generated by the computing server. TPM enhances this process by ensuring the integrity of the computation, while TEE isolates and secures execution, providing a trusted execution space that minimizes the risk of data leakage and tampering. In this method, the computing server cluster is divided into two groups: the working group is responsible for generating abstracts for articles, while the validation group verifies the abstracts generated by the working group. The cluster reaches a consensus under the allocation of a specific smart contract, ensuring the trustworthiness of the generated abstracts. The pseudocode for the above process is depicted in Algorithm 1.
\end{itemize}

The separation of file storage and computation across different servers is managed through TEE, ensuring that data operations are isolated and secure, minimizing the risk of unauthorized data leakage or tampering. P2P encrypted communication and trustworthy services jointly implement a trustworthy server, making it a reliable component of BeerReview.

\begin{algorithm}
    
    \DontPrintSemicolon  
    \SetAlgoLined 
    \caption{Summary Consensus Protocol}
    
    \SetKwInOut{Input}{Input}
    \SetKwInOut{Output}{Output}
    
    \Input{Original text}
    \Output{Trustworthy summary}
    
    \SetKwFunction{FGenerateSummary}{GenerateSummary}
    \SetKwFunction{FVerifySummary}{VerifySummary}
    \SetKwProg{Fn}{Function}{:}{\KwRet}
    \Fn{\FGenerateSummary{$text$}}{
        $model\_instance \gets \text{RandomSelect}(model\_pool)$\;
        $summary \gets model\_instance.\text{GenerateSummary}(text)$\;
        \KwRet{$summary$}\;
    }
    
    \Fn{\FVerifySummary{$summary, text$}}{
        $validators \gets \text{RandomSelect}(model\_pool, 2)$\;
        $results \gets []$\;
        \ForEach{$validator \in validators$}{
            $result \gets validator.\text{Verify}(summary, text)$\;
            $results.\text{Append}(result)$\;
        }
        \If{all $results$ are \textbf{true}}{
            \KwRet \textbf{true}\;
        }
        \KwRet \textbf{false}\;
    }
    
    $summary \gets \FGenerateSummary{\textit{original\_text}}$\;
    $is\_trustworthy \gets \FVerifySummary{summary, \textit{original\_text}}$\;
    
    \While{not $is\_trustworthy$}{
        $summary \gets \FGenerateSummary{\textit{original\_text}}$\;
        $is\_trustworthy \gets \FVerifySummary{summary, \textit{original\_text}}$\;
    }
    
    $\text{RecordInBlockchain}(summary)$\;

\end{algorithm}

\section{Implementation Details}
\subsection{Front-end UI}
\begin{figure}
    \centering
    \includegraphics[width=0.48\textwidth]{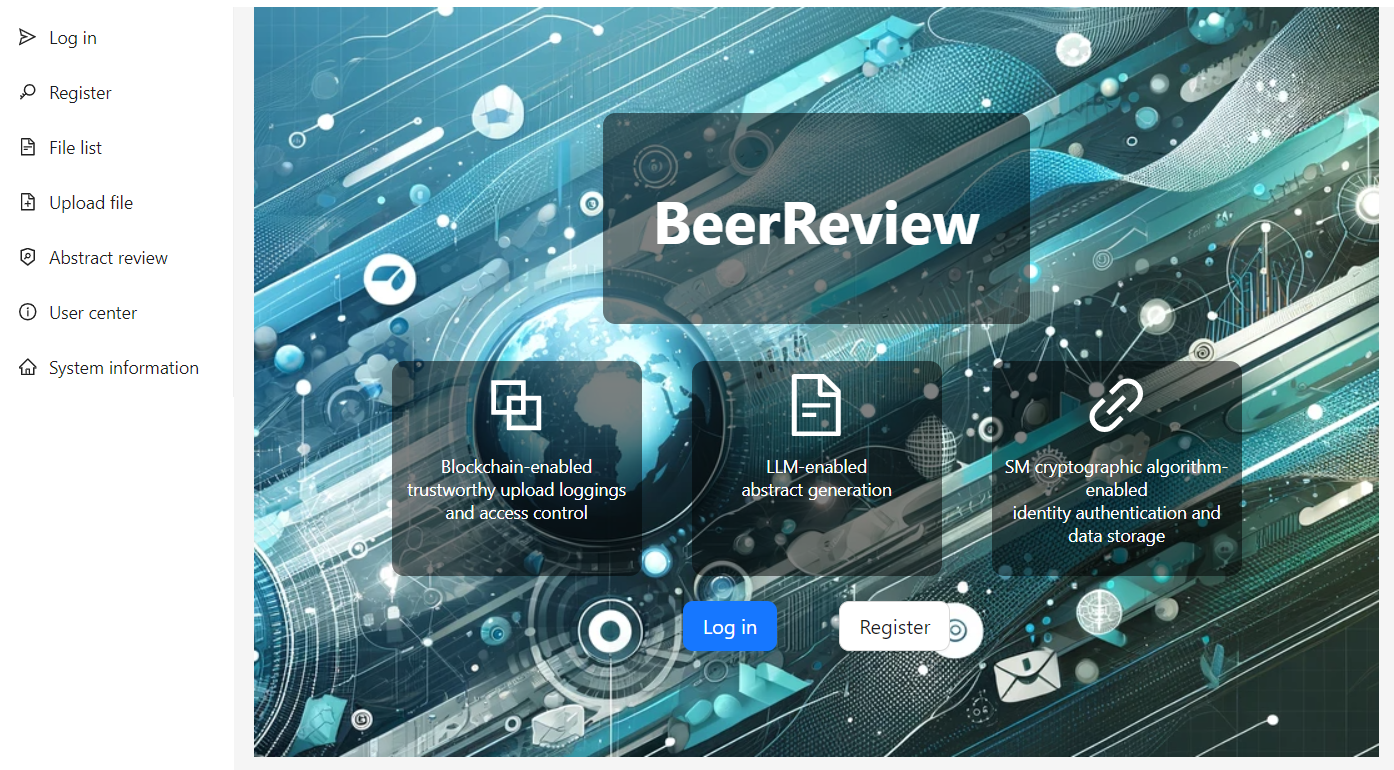}
    \caption{The Web UI of BeerReview}
    \label{fig: Web UI}
\end{figure}
The front-end UI of BeerReview is constructed on a React framework utilizing JavaScript. The web UI of BeerReview, as shown in Fig. 3, provides interactions including login, registration, file list, file uploading, abstract review, abstract review, user center, and system information.

\subsection{Blockchain Network}
The design of the blockchain network serves as a critical backbone for ensuring transparency, security, and decentralized governance. Therefore, Ethereum is chosen as the foundational blockchain framework for the BeerReview platform. Since a high transaction fee (gas) in Ethereum is required for usage, and experts and scholars shouldn't be occupied with mining to earn the fee, an ether distribution method \cite{kang2021} is adopted. In this method, the genesis account, which works as a distributor account, owns a great amount of ether. This account will send ether to a new account with a verified identity so that the user can work with the platform without worrying about the fee.

\begin{figure}
    \centering
    \includegraphics[width=0.48\textwidth]{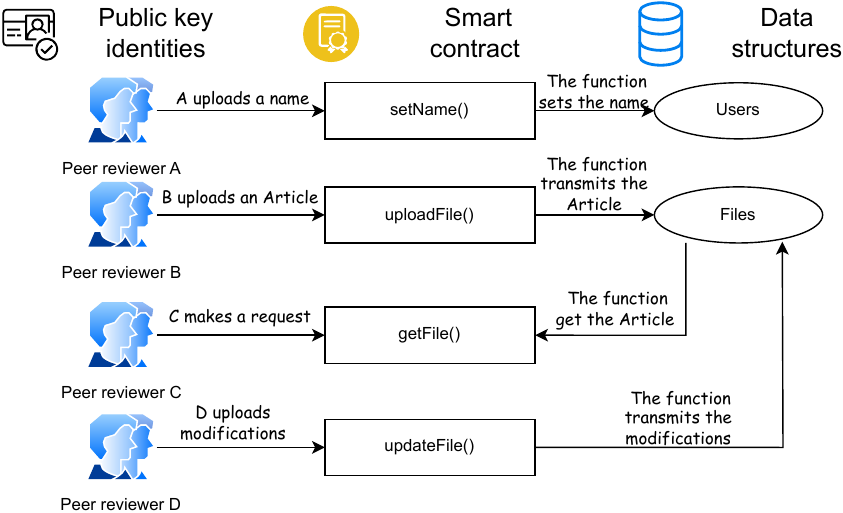}
    \caption{The design of the smart contract for BeerReview}
    \label{fig: Smart Contract}
\end{figure}

\subsection{Smart Contract}
On BeerReview, public key identities, uniquely generated from personal information, enable experts and scholars to interact with the smart contract.

The smart contract for BeerReview is shown in Fig. 4. For data structures, two main data structures are constructed for the smart contract named Users and Files for storing the data of users and Articles respectively. To operate with the data structures that are stored on Ethereum, four main functions called \textbf{\textit{setName(), uploadFile(), getFile() and updateFile()}} are written as shown in Tab. 1. The first function can create a name for a public key identity, which is shown to other users, and the second one is designed to upload information regarding the Article, such as the hash value, the uploader, the corresponding asymmetric key. The third function enables the users in the corresponding authority group to access the Article, while the final one allows the users in the authority group for the Article to update the Article. 

\begin{table}[htbp]
    \centering
    \caption{The main functions of the smart contract}
    \label{Tab: Main Functions}
    \begin{tabular}{|l|l|l|}
        \hline
        Function Name & Functionality \\
        \hline
        setName() & Creating a name \\
        uploadFile() & Uploading information of an Article \\
        getFile() & Getting an Article \\
        updateFile() & Updating modifications of an Article \\
        \hline
    \end{tabular}
\end{table}

Services are provided to the users through interaction with the smart contract for BeerReview. An interface provided by Web3j, a lightweight Java class library based on the JSON-RPC framework, is adopted to allow programs to integrate nodes on the Ethereum blockchain network \cite{guo2022}. The interface converts the smart contract to a Java class with functions by its APIs, which can be directly called by other Java classes \cite{guo2022}. It is widely discussed that web3 is an incredible tool for decentralized network systems. Here are the reasons.
\begin{itemize}
    \item Web3 brings an emerging outlook for the value of decentralization, which can be used to realize complete decentralization and to underpin services and applications, such as deController \cite{xu2023decontroller}.
    \item Furthermore, Web3 opens the world of the new existence of the crypto-network-entity, which can even be extended to the outer cyber and physical world, working as a way to construct a secure, decentralized network ecosystem \cite{xu2023web3}.
\end{itemize}

\subsection{LLM Instance}
The ChatGLM3-6B, a local LLM deployment on BeerReview, efficiently processes numerous articles by generating abstracts without requiring internet access, enhancing both efficiency and security. This advanced model from the ChatGLM series boasts a low deployment threshold and improved capabilities, supporting high-volume usage with its robust feature set \cite{zeng2022glm130b}. Due to the current limitation of only one server providing LLM instance services, trustworthy abstract consensus can't be achieved for now.

\subsection{Back-end Server}
To provide users with a better interactive experience, the web application of BeerReview is constructed on the encrypted file server, which is based on Browser/Service architecture. The BeerReview website is designed for convenient access by experts and scholars using personal computers or mobile phones, eliminating the need for tedious operations. Apache Tomcat 9.0, an open-source Java Servlet container, is deployed on the encrypted file server to provide the following functions. Please note that the subsequent discussion will be elaborated in conjunction with Fig. 2, with step 1 referred to 1 in Fig. 2, and the remaining steps similarly correspond.

\begin{itemize}
    \item \textbf{Step 1: public key identities and blockchain wallets creation} Users generate unique public-private key pairs through the SM2 algorithm during registration, enabling the creation of secure blockchain wallets via the web3j interface, with private keys held only by the users for secure platform access.
    \item \textbf{Step 3: the generation of abstracts for Articles} Upon article upload, the computing server uses a local LLM instance to produce abstracts for peer review endorsement, maintaining the integrity and confidentiality of the content.
    \item \textbf{Step 4: the computation and storage of hash values} The encrypted file server calculates the article's hash using the SM3 algorithm and stores it on Ethereum, ensuring data integrity within the decentralized network.
    \item \textbf{Step 5: Article plaintext encryption} For data protection, articles are encrypted with symmetric keys, which are then secured with corresponding public keys. The encrypted symmetric keys are stored on Ethereum, while the encrypted articles remain on the server. Decryption is exclusively controlled by the user's private key.
    \item \textbf{Step 6: experts endorsing} Experts endorsing is the core of the peer review process. A number works as the state flag is set for each Article, which has three possible values: ‘0’ for not in the peer review process, ‘1’ for in the process, and ‘2’ for having finished the process. Passing the peer review process involves certain thresholds, requiring the participation of a significant number of top experts in the field and a certain proportion of actively engaged reviewers, which can be adjusted by authoritative institutions or organizations.
    \item \textbf{Step 7-10: access to Articles and modification updates} Authorized users can read, comment on, and annotate articles. All user interactions, including modifications, are recorded on Ethereum for transparency and security.
\end{itemize}
The encrypted file server, the computing server, and Ethereum work independently with critical interactions, creating a highly decoupled back-end network system that guarantees the protection of IPRs.

\section{Alpha Test}

\begin{table*}[htbp]
    \centering
    \caption{Feature Comparison Between Different Peer Review Platforms}
    \label{Tab: Feature Comparison Between Different Peer Review Platforms}
    \begin{tabular}{|l|l|l|l|l|}
        \hline
            & Decentralization  & Interactive Review & Identity Protection & Intellectual Property Protection Requirement\\
        \hline
            BeerReview & Yes & Yes & By self-control & The trustworthy information loggings\\ 
        \hline
            ScholarOne Manuscripts \cite{manuscriptcentral} & No & No & By centralized authority & Transferring copyright before the review process\\
        \hline
            Publons \cite{publons} & No & No & By centralized authority & Transferring copyright before the review process\\
        \hline
            Samson Editorial System \cite{samsoncn} & No & No & By centralized authority & Transferring copyright before the review process \\
        \hline
            Magtech Editorial System \cite{magtech} & No & No & By centralized authority & Transferring copyright before the review process \\
        \hline    
    \end{tabular}
\end{table*}

The preliminary alpha test for the BeerReview system was designed to evaluate its basic functionalities and workflow efficiency. The infrastructure supporting BeerReview includes two AWS cloud servers and a PC for the LLM instance. The blockchain server, equipped with a 2.5GHz CPU and 4GB of RAM, hosts an Ethereum node and manages a private blockchain based on PoS consensus. The backend server, which has a 2.5GHz CPU and 8GB of RAM, handles application logic and user interactions. The PC for the LLM instance, vital for processing natural language tasks, features an i5-11500 CPU, an RTX 3060 GPU, and 16GB of RAM. It manages to generate article abstracts at an average speed of 30 seconds per abstract, balancing speed and accuracy to enhance the review process efficiency.

During this initial phase, BeerReview successfully processed 19 articles, 31 comments, and 49 contributed by 23 users, accumulating a total word count of 54,701. This test served primarily to verify the system’s functionality and to refine the operational procedures. Feedback from users provided crucial insights into the system's performance and highlighted areas for further enhancement. Comparative feature analysis with established peer review platforms such as ScholarOne Manuscripts, Publons, Samson Editorial System, and Magtech Editorial System was conducted, focusing on decentralization, tamper resistance, transparency, process speed, and intellectual property protection, as shown in Tab. 2.

\section{Conclusion}
The paper has demonstrated the design and implementation details of BeerReview. The alpha test for BeerReview has been finished to facilitate experts and scholars to conduct the peer review process and protect their IPRs. Feedback from users indicates that BeerReview has been positively received by scholars and experts who have used the platform. The platform's effective measures in protecting intellectual property rights and preventing information tampering suggest its potential applicability to other scenarios, such as note-taking systems. Future aspirations include the adoption of BeerReview by educational institutions to further enhance the efficacy of peer review systems and reinforce intellectual property protections.

\bibliographystyle{IEEEtran}
\bibliography{main}

\vfill

\end{document}